\documentclass[aps,pra,twocolumn,superscriptaddress,notitlepage,nofootinbib,longbibliography, colorlinks=true]{revtex4-2}
\usepackage{amsmath,amssymb,amsfonts,graphicx,float,times,psfrag}
\usepackage[pdftex]{color}
\usepackage{amsmath,bm}
\usepackage[colorlinks, linkcolor=blue, citecolor=blue,  breaklinks=true]{hyperref}
\usepackage{mathtools,amsfonts,mathptmx}
\usepackage[utf8]{inputenc}
\usepackage[T1]{fontenc}
\usepackage{xcolor}
\usepackage{braket}
\usepackage[titletoc,title]{appendix}
\usepackage[nameinlink,capitalize]{cleveref}

\usepackage[shortlabels]{enumitem}
\begin{document}

\title{Quantum entanglement enhanced via dark mode control in  molecular optomechanics}

\author{E. Kongkui Berinyuy}
\email{emale.kongkui@facsciences-uy1.cm}
\affiliation{Department of Physics, Faculty of Science, University of Yaounde I, P.O.Box 812, Yaounde, Cameroon}

\author{P. Djorwé}
\email{djorwepp@gmail.com}
\affiliation{Department of Physics, Faculty of Science, 
University of Ngaoundere, P.O. Box 454, Ngaoundere, Cameroon}
\affiliation{Stellenbosch Institute for Advanced Study (STIAS), Wallenberg Research Centre at Stellenbosch University, Stellenbosch 7600, South Africa}

 \author{A. N. Al-Ahmadi}
 \email{anahmadi@uqu.edu.sa}
 \affiliation{Department of Physics, College of Sciences, Umm Al-Qura University, Makkah 24382, Saudi Arabia}

 \author{H. Ardah}
 \email{hyabdullrahman@pnu.edu.sa}
 \affiliation{Department of Computer Sciences, College of Computer and Information Sciences, Princess Nourah bint Abdulrahman University, P.O.Box 84428, Riyadh 11671, Saudi Arabia
 	} 

\author{A.-H. Abdel-Aty}
\email{amabdelaty@ub.edu.sa}
\affiliation{Department of Physics, College of Sciences, University of Bisha, Bisha 61922, Saudi Arabia}


\begin{abstract}
Quantum entanglement is an interesting resource for modern quantum technologies, where generating multiple quantum entanglement is highly required. However, entanglement engineering between multiple modes
is strongly suppressed by dark mode effect. Here, we proposed a scheme based on molecular cavity optomechanical structure that enhances quantum bipartite and tripartite entanglement via dark mode breaking. Our proposal consists of an optical cavity that hosts two molecular ensembles which are coupled through an intermolecular coupling. A vibrational hopping rate $J_m$ captures the intermolecular coupling that is phase modulated via the synthetic gauge field method. The breaking of the dark mode is controlled by tuning both the intermolecular coupling and its modulation phase. By adjusting these parameters in our proposal, we can flexibly switch between the Dark Mode Unbroken (DMU) and the Dark Mode Broken (DMB) regimes. We find that
in the dark-mode-unbroken regime, the amount of the generated bipartite and tripartite entanglement is significantly low or is suppressed. In contrast, in the dark-mode-broken regime, the entanglement is greatly enhanced,i.e., up to twofold enhancement. Moreover, the generated entanglement is more resilient against thermal noise in the dark-mode-broken regime compared to the thermal robustness in the unbroken regime. Therefore, our proposed scheme serves as a benckmark system to improve quantum correlations engineering, and to generate noise-tolerant quantum resources for applications in numerous modern quantum technologies.
\end{abstract}

\maketitle
\date{\today}

\section{Introduction}\label{intro}
Molecular cavity optomechanical structures involving
molecular ensembles have attracted increasing attention owing to physical phenomena beyond those achievable in conventional cavity optomechanical systems~\cite{Gu2021,Thomas2016,Neuman2018,Berinyu2025,Peng2025}. In molecular cavity optomechanical systems, confined electromagnetic
fields couple strongly to molecular vibrational modes, allowing coherent control of molecular motion at the quantum level. This capability makes molecular cavity optomechanics a powerful platform for exploring quantum effects, including quantum entanglement~\cite{Lombard,Huang2024,Berinyuy2025,BERINYUY2025417313,BERINYUY2026117820}. 

  Quantum entanglement, a fundamental pillar of quantum
  mechanics~\cite{Horodecki2009} is characterized by intrinsically inseparable
  correlations between spatially distinct entities, and it constitutes a fundamental resource for quantum technologies such as sensing and metrology~\cite{Gerrits2021,Djo2019,Tchounda2023}, quantum information processing~\cite{Armstrong2012}, quantum computing,~\cite{Duarte2021} and secure communication~\cite{Choi2008}.  Recent experiments have successfully demonstrated and controlled such quantum correlations in cavity optomechanical systems~\cite{Vitali2007}.

In molecular cavity optomechanical systems, the generation of large-scale photon-vibrational mode entanglement remains an outstanding challenge. A central obstacle arises
from the dark-mode effect. This dark-mode is induced by coupling a single optical mode to multiple degenerates or near-degenerate ensemble of molecular modes. Consequently, collective vibrational excitations decouple from the optical
fields, thereby suppressing the effective light-vibrational motion interaction and then the generation of photon-vibrational
mode entanglement. Remarkable effort has been directed toward overcoming the dark-mode effect in cavity optomechanical systems, a critical step for enhancing light-mechanical coupling and realizing large-scale photon-phonon entanglement~\cite{Guo2023,Lai2022,Dong2012,Massem2025}. Experimental demonstration of simultaneous
cooling of two near-degenerate mechanical modes through the breaking of the mechanical dark mode in a two-membrane cavity optomechanical system was investigated in~\cite{Cao2025}. Overcoming dark-mode suppression is crucial as it induces a generation of large amount of quantum correlations which are
useful in modern quantum networks~\cite{Kimble2008,Mastriani}. Furthermore, dark-modes breaking can serve to shield systems from mechanical dissipation~\cite{Dong2012}, facilitate high-fidelity transduction of quantum state and enable efficient routing and switching of photons across different wavelengths~\cite{Lai2022,Lake,Tian2012}. Despite the aforementioned potential applications enabled by breaking dark modes in conventional optomechanical systems, darkmode breaking in molecular cavity optomechanics remains largely unexplored and constitutes the central focus of this work.

Here, we proposed a theoretical scheme to generate tripartite entanglement by breaking dark-mode in a molecular cavity optomechanical system. The breaking of the dark-mode in our proposal is realized by tuning a synthetic magnetism induced via the inter-molecular interaction. We find that in the Dark-Mode-Unbroken regime (DMU), the amount both
bipartite and tripartite entanglement is weak or suppressed. In contrast, in the Dark-Mode-Broken (DMB) regime, the generated entanglements are greatly enhanced. This shows that the
intermolecular coupling $J_m$ plays a key role in controlling and engineering entanglement in our proposed scheme. Indeed,
within the DMU regime, the dark mode remains effectively
decoupled from the cavity field. As a result, it acts as a reservoir for excitations and prevents efficient energy and quantum
correlation exchange between the cavity mode and the molecular ensemble. This weak interaction suppresses quantum correlations among the subsystems, leading to weak amount of
both bipartite and tripartite entanglement. In the DMB regime however, the dark mode is activated through dark-mode breaking. This activation allows the previously isolated mode to hybridize with both the cavity and molecular modes, opening
new interaction channels. The enhanced mode coupling facilitates energy transfer and strengthens quantum correlations across the system, significantly enhancing both bipartite and tripartite entanglement. Therefore, by tuning the intermolecular coupling $J_m$ one can control the transition between the DMU and DMB regimes, effectively switching entanglement off or on. This tunability explains why $J_m$ plays a crucial role
in entanglement engineering in the proposed scheme.

The rest of work is organized as follow. \Cref{sec:model} introduces the theoretical model and outlines the derivation of the dynamical equations. \Cref{sec:DM} delves into the analyzing of the dark mode control and numerical results. \Cref{sec:concl} provides concluding remarks.

\section{Model and dynamical equations} \label{sec:model}
\begin{figure}[htp!]
	\setlength{\lineskip}{0pt}
	\centering
	\includegraphics[width=.95\linewidth]{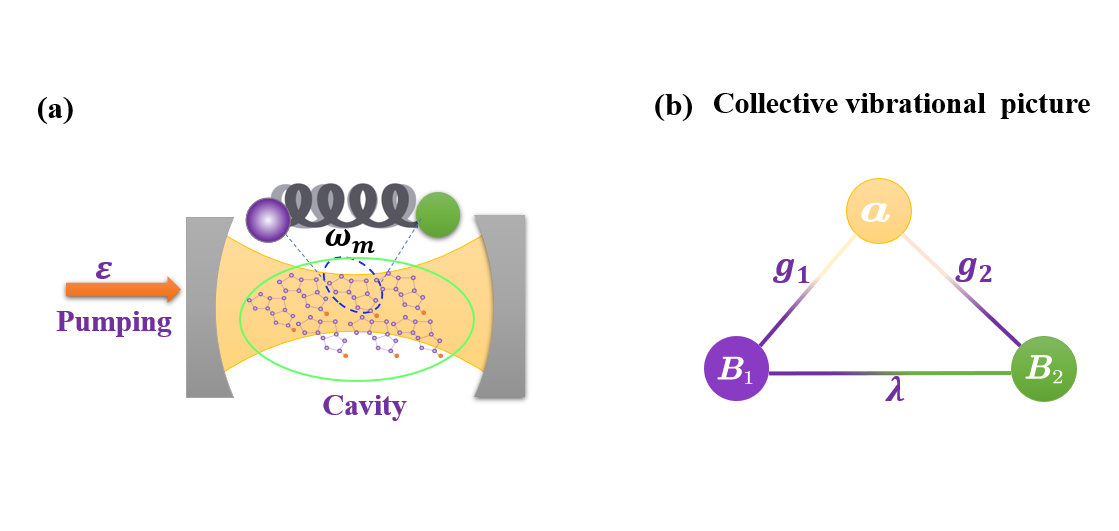}
    \caption{Sketch of our benckmark system. (a) A molecular cavity optomechanical system, illustrating the interaction between the intracavity mode and the ensemble of $N$ identical molecules with a vibrational frequency $\omega_{m}$. The vibrational mode is depicted as two masses
   connected by a spring, emphasizing the coupling between molecular vibrations. (b) Diagram illustrating the interactions between the subsystems, the collective vibrational modes of the molecules and the intracavity mode.}
	\label{fig:Setup}
\end{figure}

Our benckmark system consists on an optomechanical cavity hosting different type of molecular ensembles, which are coupled through an intermolecular rate $J_m$. The intermolecular coupling involves a phase modulation that is similar to a synthetic magnetism jauge field found in optomechanical structures \cite{S2021,Song2025}.  A strong driving field, with amplitude $\mathcal{E}$ and frequency $\omega_{l}$ is applied to drive the cavity hosting the molecular ensembles as depicted in \Cref{fig:Setup}.

The molecular vibration can be model as a quantum harmonic oscillator.
The Hamiltonian of the system in question reads ($\hbar$=1),
\begin{equation}\label{eq:1}
\begin{aligned}
\mathcal{H}=&\Delta_a a^\dagger a+\sum_{j=1}^N \omega_m b_j^\dagger b_j+\sum_{j=1}^N g_m a^\dagger a(b_j^\dagger + b_j)\\&+J_m\sum_{j=1}^{N-1}(e^{i\theta}b_j^\dagger b_{j+1}+e^{-i\theta}b_{j+1}^\dagger b_j)+ \mathcal{E}(a+a^\dagger),
\end{aligned}
\end{equation}
where $a (a^\dagger)$ and $b_j (b_j^\dagger)$ are the annihilation (creation) operators for the cavity and mechanical modes (with $j = 1, 2$), obeying the commutation relations $[\mathcal{O}_j, \mathcal{O}_j^\dagger] = 1$, $\mathcal{O}=a,b_j$. The resonance frequency of the vibrational mode is $\omega_m$. The quantity $\Delta_{a}=\omega_{a}-\omega_{l}$ in the above Hamiltonian represents the frequency detuning that is introduced by moving in the frame rotating at the driving frequency $\omega_{l}$, where $\omega_{a}$ is the resonance frequency of the cavity. The first two terms of our Hamiltonian capture the free energies of the plasmonic cavity and for the molecular vibrational modes. The fourth term represents the optomechanical coupling between the intracavity mode and the molecular vibrational mode through the coupling strength $g_m$. The last term captures the driving field.

 We will consider a large number of molecules in the cavity, so that we can introduce the molecular collective operators,
\begin{equation}\label{eq:3}
B_1=\sum_{j=1}^{M}b_j/\sqrt{M}, ~~~B_2=\sum_{j=M+1}^{N}b_j/\sqrt{N-M}.
\end{equation}
\\

This separation into two collective modes, $B_1$ and $B_2$, is a theoretical construction that allows for simplifying the analysis and does not imply that the molecules are physically separated into two distinct groups. With these modes, Our Hamiltonian takes the form,
\begin{equation}\label{eq:4}
\begin{aligned}
\mathcal{H}=&\Delta_a a^\dagger a+\sum_{k=1}^2\left(\omega_k B_k^\dagger B_k+g_k a^\dagger a(B_k^\dagger + B_k)\right)+\lambda(e^{i\theta}B_1^\dagger B_2\\&+e^{-i\theta}B_2^\dagger B_1)+\mathcal{E}(a+a^\dagger)
\end{aligned}
\end{equation}
where $g_1=g_m\sqrt{M}$ and $g_2=g_m\sqrt{N-M}$ represent the collective optomechanical coupling strength between the cavity mode and the collective vibrational mode $B_1(B_2)$. The quantities $M$ and $N$ are the distribution number of molecular collective mode and number of molecules, respectively. In the Hamiltonian above, $\lambda$ is collective optomechanical coupling strength between the two collective modes defined by $\lambda=J_m\sqrt{M(N-M)}$.

\subsection{\label{sec:IIB}Quantum Langevin equations (QLE)} 

The dynamical state of the system under consideration is described by the following Quantum Langevin Equations (QLEs)
\begin{equation}\label{eq:5}
\begin{aligned}
\dot{a}=&-(i\Delta_{a}+\kappa)a-ia\sum_{k=1}^{2}g_k(B_k+B_k^\dagger)-i\mathcal{E}+\sqrt{2\kappa}a^{\text{in}},\\
\dot{B}_1=&-(i\omega_m+\gamma_1)B_1-ig_1a^\dagger a-i\lambda e^{i\theta}B_2+\sqrt{2\gamma_1}B_1^{\text{in}},\\
\dot{B}_2=&-(i\omega_m+\gamma_2)B_2-ig_2a^\dagger a-i\lambda e^{-i\theta}B_1+\sqrt{2\gamma_2}B_2^{\text{in}},
\end{aligned}
\end{equation} 
where the related noise operators $\mathcal{O}^{\text{in}} (\mathcal{O}=a,B_j)$ have been taken into consideration. At this juncture, we would like to mention that,
\begin{align}
&B^{\text{in}}_1=\sum_{j=1}^{M}b^{\text{in}}_j/\sqrt{M},
&B^{\text{in}}_2=\sum_{j=M+1}^{N}b^{\text{in}}_j/\sqrt{N-M}.
\end{align} 
These noise operators have zero mean values, and are characterized by the correlation functions,
\begin{equation}\label{eq:6}
\begin{aligned}
&\langle a^{\text{in}}(t)a^{\text{in}\dagger}(t^\prime)\rangle=\delta(t-t^\prime),\\
&\langle B_k^{\text{in}}(t)B_k^{\text{in}\dagger}(t^\prime)\rangle=(n_k+1)\delta(t-t^\prime),\\
&\langle B_k^{\text{in}\dagger}(t)B_k^{\text{in}}(t^\prime)\rangle=n_k\delta(t-t^\prime),
\end{aligned}
\end{equation}
where $n_j=\left[\exp\left(\frac{\hbar\omega_j}{k_B\text{T}}\right)-1\right]^{-1}$ denotes the thermal phonon number at temperature $T$ and $k_B$ is the Boltzmann constant. We assume that $\gamma_1$=$\gamma_2$=$\gamma$ for the two collective vibrational modes. For the cavity modes, the thermal excitation number can be neglected in the optical frequency band. A linearization procedure can be used to simplify the physical model by considering a strong enough intracavity population driving the molecular ensembles. Following this procedure, each operator ($\mathcal{O}$) is split into their mean value ($\langle\mathcal{O}\rangle$) plus small fluctuations ($\delta\mathcal{O}$) around them, i.e., $\mathcal{O}=\langle\mathcal{O}\rangle+\delta\mathcal{O}$, where $\mathcal{O}\equiv a,B_k$, and $\langle\mathcal{O}\rangle\equiv\alpha, \beta_k$. For long term behavior, the mean value equations are no longer time dependent, and they reach the following steady state equations,
\begin{equation}
\begin{aligned}
\dot{\alpha}=&-(i\tilde{\Delta}_a+\kappa)\alpha-i\mathcal{E},\\
\dot{\beta_1}=&-(i\omega_m+\gamma_1)\beta_1-ig_1|\alpha|^2-i\lambda e^{i\theta}\beta_2,\\
\dot{\beta_2}=&-(i\omega_m+\gamma_2)\beta_2-ig_2|\alpha|^2-i\lambda e^{-i\theta}\beta_1.
\end{aligned}
\end{equation}
Similarly, the set of equations for the fluctuation operators read,
\begin{equation}\label{eq:7}
\begin{aligned}
\delta\dot{a}=&-(i\tilde{\Delta}_a+\kappa)\delta a-i\sum_{j=1}^{2}G_j(\delta B_j+\delta B_j^\dagger)+\sqrt{2\kappa}a^{\text{in}},\\
\delta\dot{B}_1=&-(i\omega_m+\gamma_1)\delta B_1-iG^\ast_1\delta a-iG_1\delta a^\dagger-i\lambda e^{i\theta}\delta B_2+\sqrt{2\gamma_1}B_1^{\text{in}},\\
\delta\dot{B}_2=&-(i\omega_m+\gamma_2)\delta B_2-iG^\ast_2\delta a-iG_2\delta a^\dagger-i\lambda e^{-i\theta}\delta B_1+\sqrt{2\gamma_2}B_2^{\text{in}},
\end{aligned}
\end{equation}
where $\tilde{\Delta}_a=\Delta_a+2\sum_{l=1,2}g_l\text{Re}[\beta_l]$ is the effective detuning, $G_1=\sqrt{M}g_m\alpha$, and $G_2=\sqrt{N-M}g_m\alpha$, are the effective optomechanical coupling strengths.

In order to study the entanglement, we define the following quadrature operators, 
\begin{equation}
\begin{aligned}
\delta x&=\frac{( \delta a+\delta a^{\dagger})}{\sqrt{2}}, \delta y=\frac{( \delta a- \delta a^{\dagger})}{i\sqrt{2}},\\
\delta Q_k&=\frac{(\delta B_k+\delta B_k^{\dagger})}{\sqrt{2}},\delta P_k=\frac{(\delta B_k-\delta B_k^{\dagger})}{i\sqrt{2}},
\end{aligned}
\end{equation}
with $k=1,2$. The corresponding quadrature  noise operators are,
\begin{equation}
\begin{aligned}
x^{\text{in}}&=\frac{(a^{\text{in}}+ a^{\text{\text{in}}\dagger})}{\sqrt{2}},~ y^{\text{in}}=\frac{( a^{\text{\text{in}}}- a^{in\dagger})}{i\sqrt{2}},\\
Q^{\text{in}}_k&=\frac{( B_k^{\text{in}}+ B_k^{\text{in}\dagger})}{\sqrt{2}},~ P^{\text in}_k=\frac{( B_k^{\text{in}}- B_k^{in\dagger})}{i\sqrt{2}}.
\end{aligned}
\end{equation}
By using these quadratures operators, our linearized equations displayed in \Cref{eq:7} can be written in the following compact form,
\begin{equation}\label{eq:8}
\dot{\Gamma}(t)=A\Gamma(t)+n(t),
\end{equation} 
where A is $6\times 6$ matrix which reads,
\begin{widetext}
\begin{equation}
A=
\begin{pmatrix}
-\kappa&\tilde{\Delta}_a&2\text{Im}[G_1]&0&2\text{Im}[G_2]&0\\
-\tilde{\Delta}_a&-\kappa&-2\text{Re}[G_1]&0&-2\text{Re}[G_2]&0\\
0&0&-\gamma_1&\omega_{m}&\lambda\sin\theta&\lambda\cos\theta\\
-2\text{Re}[G_1]&-2\text{Im}[G_1]&-\omega_{m}&-\gamma_1&-\lambda\cos\theta&\lambda\sin\theta\\
0&0&-\lambda\sin\theta&\lambda\cos\theta&-\gamma_2&\omega_{m}\\
-2\text{Re}[G_2]&-2\text{Im}[G_2]&-\lambda\cos\theta&-\lambda\sin\theta&-\omega_{m}&-\gamma_2\\
\end{pmatrix},
\end{equation}
\end{widetext}
with the vector $\Gamma^\top$ defined as,
	\begin{equation}
	\Gamma^{\top}(t)=\left(\delta x(t),\delta y(t),\delta Q_{1}(t),\delta P_{1}(t),\delta Q_2(t),\delta P_2(t)\right),
	\end{equation} and the fluctuation operator $n^{\top}$ that reads,
	\begin{equation}
	n^{\top}(t)=\left(\sqrt{2\kappa}x^{\text{in}},\sqrt{2\kappa}y^{\text{in}},\sqrt{2\gamma_1}Q_{1}^{\text{in}},\sqrt{2\gamma_1}P_{1}^{\text{in}},\sqrt{2\gamma_2}Q_2^{\text{in}}, \sqrt{2\gamma_2}P_2^{\text{in}}\right).
	\end{equation}
	
The system under consideration is stable, meaning that all the real parts of the eigenvalues of drift matrix $A$ are negative. This stability conditions can be derived by using Routh-Hurwitz criterion \cite{Dejesus1987}. Owing to the Gaussian nature of quantum noises and the linearity of the QLEs, the system can fully be characterised by $ 6\times6$ covariance matrix (CM) which can be obtain by solving the following Lyapunov equation,
\begin{equation}
AV+VA^{\top}=-D,
\end{equation}
where $V_{lk}=\frac{1}{2}\left\{\Gamma_{l}(t)\Gamma_{k}(t^{\prime})+\Gamma_{k}(t^{\prime})\Gamma_{l}(t)\right\}$, the diffusion matrix is given by, 
\begin{equation}
D=\text{diag}\left[\kappa,\kappa,\gamma_1(2n_{\text{th}}+1),\gamma_1(2n_{\text{th}}+1),\gamma_2(2n_{\text{th}}+1),\gamma_2(2n_{\text{th}}+1)\right].
\end{equation}

\subsection{Quantification of bipartite entanglement}\label{IIC}
In order to quantify entanglement in our system, we use logarithmic negativity $E_{n}$ as a measure to evaluate bipartite entanglement while the residual minimum contangle $\mathcal{R}_{\tau}^{min}$ is used to quantify genuine tripartite entanglement. The mathematical expression for this logarithmic negativity is given by~\cite{Plenio2005},
\begin{equation}
E_{n}=\text{max}\left[0,-\text{ln}(2\nu)\right],
\end{equation}
where $\nu\equiv2^{-1/2}\left\{\sum(V)-\left[\sum(V)^{2}-4\text{det}(V)\right]^{1/2}\right\}^{1/2}$ with $\sum(V)=\text{det}(\Psi_{1})+\text{det}(\Psi_{2})-2\text{det}(\Psi_{3})$. 
The $\Psi_{j}$ are the elements of the considered bipartite submatrix extracted from the covariance matrix of the whole system. For a given subsystem consisting of mode $1$ and $2$, the related submatrix reads,
\begin{equation}
V_{sub}=
\begin{pmatrix}
\Psi_{1}&\Psi_{3}\\
\Psi_{3}^{T}&\Psi_{2}\\
\end{pmatrix}.
\end{equation}

\subsection{Quantification of tripartite entanglement}
The minimum residual contangle $\mathcal{R}_{\tau}^{min}$ is used to quantify tripartite quantum entanglement. It is define as, 
\begin{equation}
\mathcal{R}_{\tau}^{min}\equiv \text{min}\left[\mathcal{R}_{\tau}^{a|B_1B_2},\mathcal{R}_{\tau}^{B_1|aB_2},\mathcal{R}_{\tau}^{B_2|aB_1}\right].
\end{equation}
This expression guarantees the invariance of tripartite entanglement under all possible permutations of the modes $\mathcal{R}_{\tau}^{r|st}\equiv \mathcal{C}_{r|st}-\mathcal{C}_{r|s}-\mathcal{C}_{r|t},~(r,s,t=a,B_1,B_2)$ satisfying the monogamy of quantum entanglement $\mathcal{R}_{\tau}^{r|st}\geq 0$.
Here, $\mathcal{C}_{u|v}$ is the contangle of subsystem $u$ and $v$ ($v$ contains one or two modes), which can be defined as squared logarithmic negativity and it is a proper entanglement monotone~\cite{Plenio2005} i.e., $\mathcal{C}_{u|v}=E^2_{N_{u|v}}$, with 
\begin{equation}
E_{N}\equiv \text{max}\left[0,-\ln2\zeta\right],
\end{equation}
where 
\begin{equation}
\zeta=\text{min}~ \text{eig}[i\Omega_{3}\zeta_{6}^\prime],
\end{equation}
with $\Omega_{3}$ and $\zeta_{6}^\prime$ defined respectively as,
 \begin{equation}
\Omega_{3}=\bigoplus_{k=1}^3i\sigma_{y},~~ \sigma_{y}=\begin{pmatrix}
0&-i\\i&0
\end{pmatrix}
\end{equation} 
and 
\begin{equation}
\zeta^\prime_{6}=P_{r|st}V_{6}P_{r|st}~~ \text{for}~~ r,s,t=a,B_1,B_2
\end{equation}
where 
\begin{equation}
\begin{aligned}
P_{a|B_1B_2}&=\text{diag(1,-1,1,1,1,1)},\\ P_{B_1|aB_2}&=\text{diag(1,1,1,-1,1,1)},\\ P_{B_2|aB_1}&=\text{diag(1,1,1,1,1,-1)},
\end{aligned}
\end{equation}

are partial transposition matrices and $V_6$ is $6\times 6$ CM of the three modes.  

\section{Dark mode breaking enhancing entanglement}\label{sec:DM}
This section presents our results and the related discussion. As stated before, both bipartite and tripartite entanglement generated in our investigation are significantly enhanced within the Dark Mode Breaking (DMB) regime. To unveil the dark mode breaking conditions, we consider our linearized Hamiltonian,
 \begin{equation}
 \begin{aligned}
     H_{lin}=& \tilde{\Delta}^{'}\delta\alpha^\dagger\delta\alpha + \sum_{j=1,2}\left[\omega_j^{'}\delta \beta_j^\dagger\delta \beta_j +G_j (\delta\alpha^\dagger\delta\beta_j + \delta\alpha\delta\beta_j^\dagger )\right]  \\&+\lambda(e^{i\theta}\delta \beta_1^\dagger\delta \beta_2 + e^{-i\theta}\delta \beta_1\delta \beta_2^\dagger).
     \end{aligned}
 \end{equation}

For $\lambda=0$, we introduce the following two mechanical bright ($B_+$) and dark ($B_-$) modes, i.e., $B_{\pm}=(G_{1(2)}\delta \beta_1 \pm G_{2(1)}\delta \beta_2)/G$, with $G=\sqrt{G_1^2 + G_2^2}$. By using these modes back in the linearized Hamiltonian leads to,
\begin{equation}
\begin{aligned}
    H_{lin}= &\tilde{\Delta}^{'}\delta\alpha^\dagger\delta\alpha + \sum_{k=\pm}\omega_k B_k^\dagger B_k +G_+ (\delta\alpha B_+^\dagger +\delta\alpha^\dagger B_+) \\&+ G_-(B_+^\dagger B_- + B_-^\dagger B_+),
\end{aligned}
 \end{equation}
where $\omega_{\pm}=\frac{\omega_1^{'}G_{1(2)}^2+\omega_2^{'}G_{2(1)}^2}{G^2}$, $G_+=G$, and $G_-=\frac{G_1G_2(\omega_1^{'}-\omega_2^{'})}{G}$. It can be seen that for $\omega_1^{'}=\omega_2^{'}$, one has $G_-=0$ and the dark mode $B_-$ is decoupled from the system, and that suppress any expectation of tripartite entanglement in our system. To expect tripartite entanglement, we consider $\lambda\neq 0$ and use the following new bosonic modes, $\tilde{B}_{\pm}=f\delta\beta_{1(2)}\mp e^{\pm i\theta}h\delta\beta_{2(1)}$, which involve the phase-dependent vibration coupling  $\theta$.  By using these expressions into the linearized Hamiltonian, one gets,
\begin{equation}
\begin{aligned}
    H_{lin}=&\tilde{\Delta}^{'}\delta\alpha^\dagger\delta\alpha + \sum_{k=\pm}\tilde{\omega}_k^{'} \tilde{B}_k^\dagger \tilde{B}_k + \tilde{G}_+ (\delta\alpha \tilde{B}_+^\dagger +\delta\alpha^\dagger \tilde{B}_+) +\\& \tilde{G}_-(\delta\alpha \tilde{B}_-^\dagger +\delta\alpha^\dagger \tilde{B}_-),
\end{aligned}
 \end{equation}
where we have defined $\tilde{\omega}_{\pm}^{'}=(\omega_1^{'}+\omega_2^{'} \pm \sqrt{(\omega_1^{'}+\omega_2^{'})^2 + 4\lambda^2})/2$, and $\tilde{G}_{\pm}=fG_{1(2)} \mp  e^{\mp i\theta}hG_{2(1)}$. The coefficients $f$ and $h$ are defined as $f=\frac{|\tilde{\omega}_{-}^{'}-\omega_{1}^{'}|}{\sqrt{(\tilde{\omega}_{-}^{'}-\omega_{1}^{'})^2 +\lambda^2}}$, and $h=\frac{f \lambda}{\tilde{\omega}_{-}^{'}-\omega_{1}^{'}}$. By considering that our molecules ensemble are degenerated, i.e., $\omega_1^{'}=\omega_2^{'}\equiv \omega_m$, one can show that $\tilde{\omega}_{\pm}^{'}=\omega_m \pm \lambda$, $f=\frac{1}{\sqrt{2}}$, $h=-\frac{1}{\sqrt{2}}$, and therefore $\tilde{G}_{\pm}=\frac{1}{\sqrt{2}}(G_{1(2)} \pm  e^{\mp i\theta}G_{2(1)})$. Moreover, if we assume the same effective optomechanical coupling for each molecule ensemble ($G_1=G_2\equiv G_m$), one gets $\tilde{G}_{\pm}=\frac{G_m}{\sqrt{2}}(1 \pm  e^{\mp i\theta})$.
\begin{figure}[tbh]
\begin{center}
  \resizebox{0.45\textwidth}{!}{
  \includegraphics{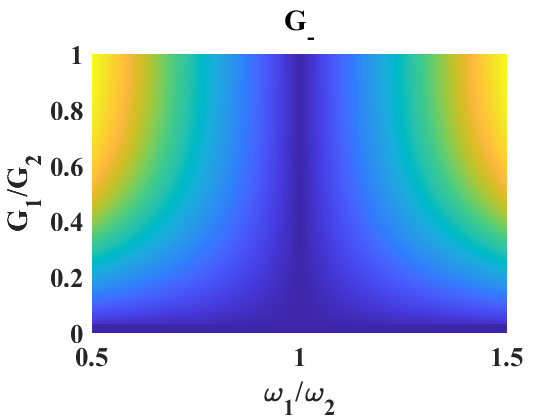}}
  \resizebox{0.45\textwidth}{!}{
  \includegraphics{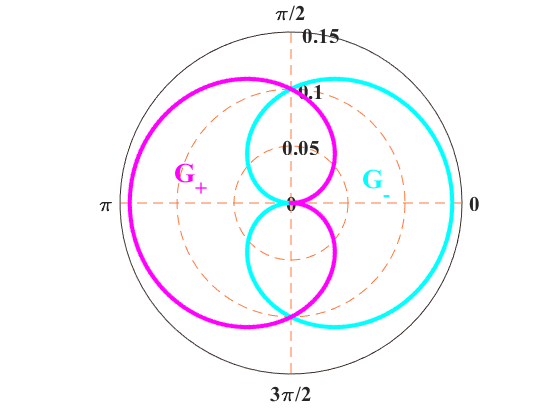}}
  \end{center}
\caption{(a) Coupling strength $G_-$ versus the coupling ratio $G_1/G_2$ and the vibration frequency ratio $\omega_1/\omega_2$ for $J_m=0$. (b) Polar representation for both $\tilde{G}_\pm$ for $G_1=G_2=0.1\omega_m$, and $J_m=0.1\omega_m$. The other used parameters are $\omega_1/2\pi=\omega_2/2\pi= 30THz$,  $\omega_1=\omega_2= 30THz=\omega_m$, $\kappa=1/3\omega_m$, $\gamma=10^{-4}\omega_m$, $g_v=10^{-4}\omega_m$, and $T=312 K$.}
\label{fig:fig2}
\end{figure}

\autoref{fig:fig2}a displays the coupling $G_-$ when $J_m=0$, and it can be seen that dark mode happens at $\omega_1=\omega_2$. Therefore, mismatch between these molecular ensemble frequencies may likely break the dark mode, and we stress that it may not improve tripartite entanglement. With the purpose to engineer tripartite entanglement, we consider $J_m\ne0$ and plot the couplings $\tilde{G}_\pm$ in polar representation as displayed in \autoref{fig:fig2}b. It can be seen that dark mode appears for $\theta=n\pi$ with $n$ being any integer. To break this dark mode, the phase most be taken as $\theta\neq n\pi$. In the rest of our investigation, we will consider $\theta = \pi/2$  where the two mode hybridize strongly.
 \begin{figure}[tbh]
	\begin{center}
		\resizebox{0.45\textwidth}{!}{
			\includegraphics{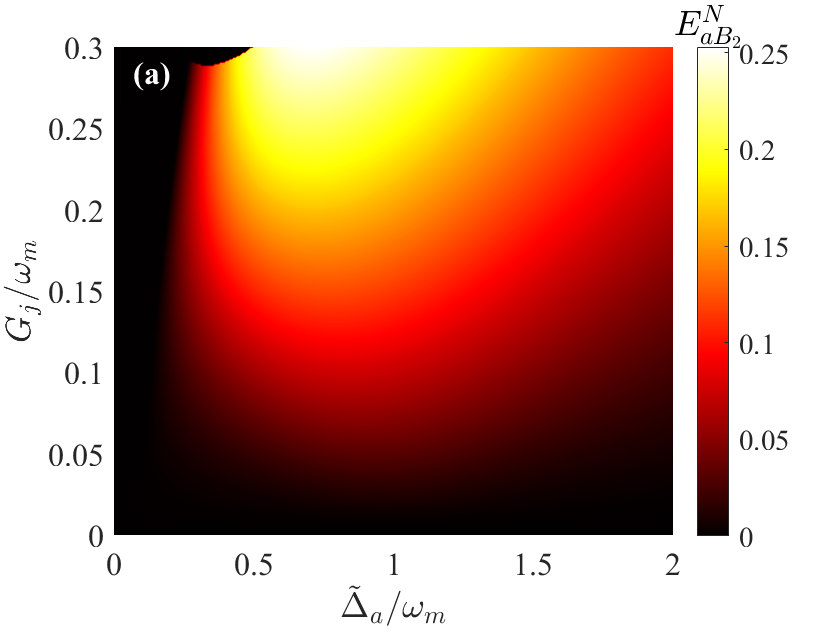}}
		\resizebox{0.45\textwidth}{!}{
			\includegraphics{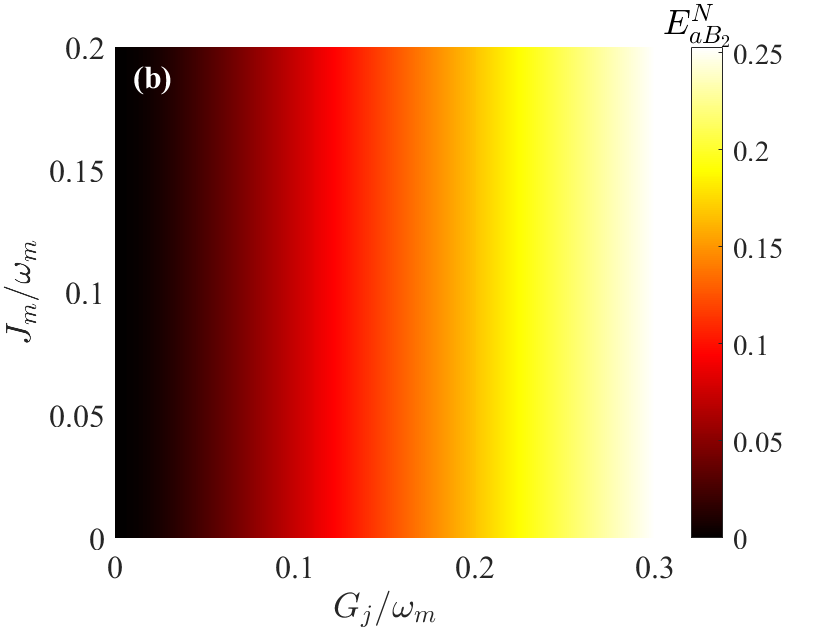}}
	\end{center}
	\caption{Logarithmic negativity (a) $E^N_{aB_2}$ vs the optomechanical coupling strength $G_j/\omega_m$ and the scaled detuning  $\tilde{\Delta}_a/\omega_m$, when $J_m/\omega_m=0$ and $\theta=0$, and  (b) $E^N_{aB_2}$ vs the optomechanical coupling strength $G_j/\omega_m$ and the coupling constant $J_m/\omega_m$, when $\tilde{\Delta}_a/\omega_m=0.7$ and $\theta=\pi/2$. The other used parameters are $\omega_1/2\pi=\omega_2/2\pi= 30THz$,  $\omega_1=\omega_2= 30THz=\omega_m$, $\kappa=1/3\omega_m$, $\gamma=10^{-4}\omega_m$, $g_v=10^{-3}\omega_m$, $T=312 K$, M=0 and N=100.}
	\label{fig:fig3}
\end{figure}

\subsection{Enhancement of mechanical bipartite entanglement}
\begin{figure*}[tbh!]
	\begin{center}
		\resizebox{0.45\textwidth}{!}{
			\includegraphics{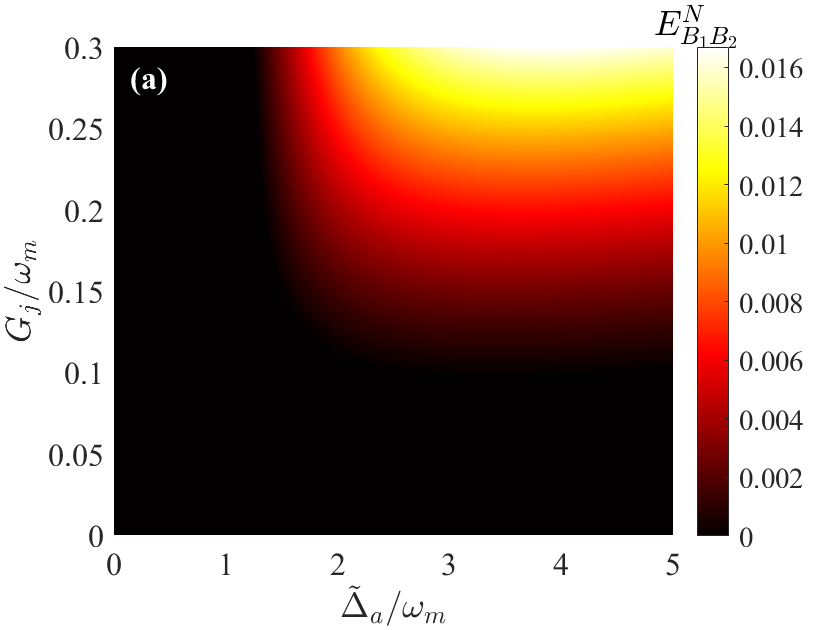}}
		\resizebox{0.45\textwidth}{!}{
			\includegraphics{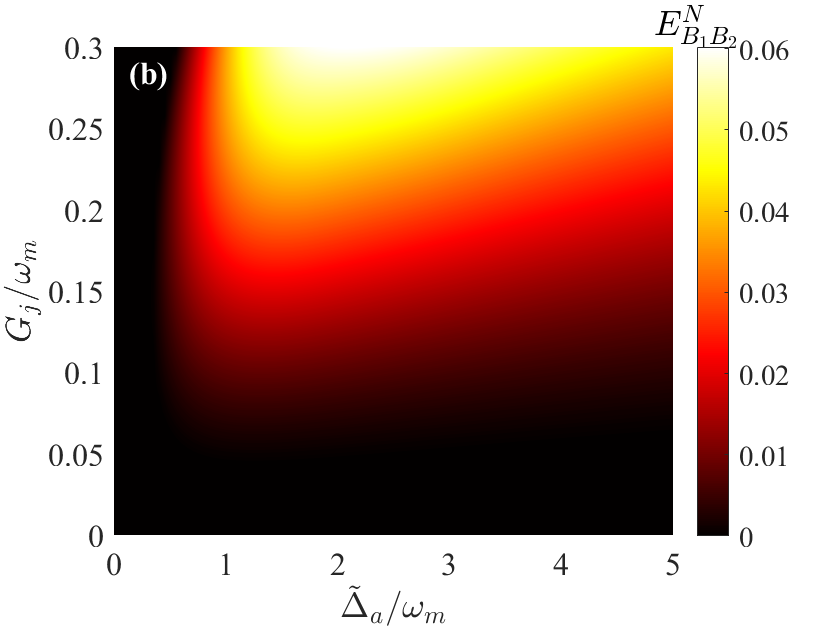}}
		\resizebox{0.45\textwidth}{!}{
			\includegraphics{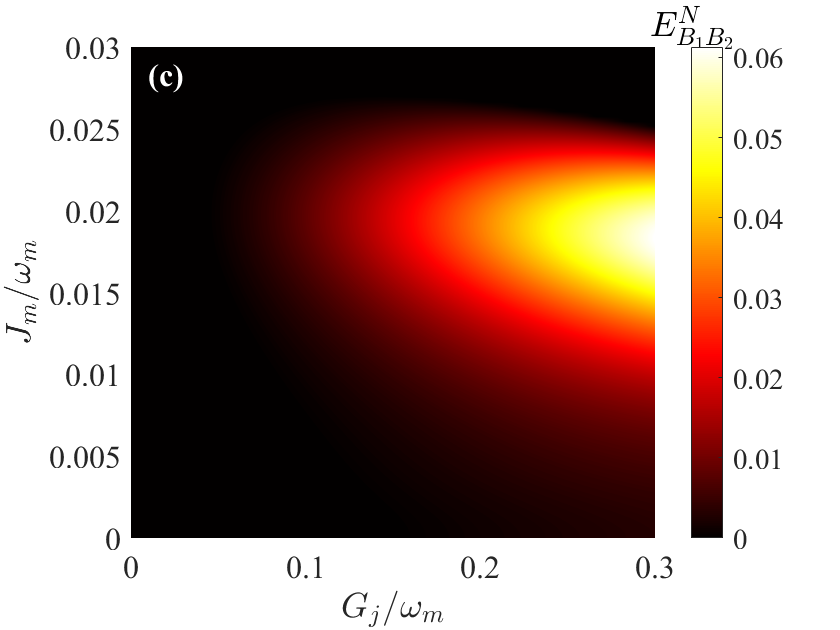}}
		\resizebox{0.45\textwidth}{!}{
			\includegraphics{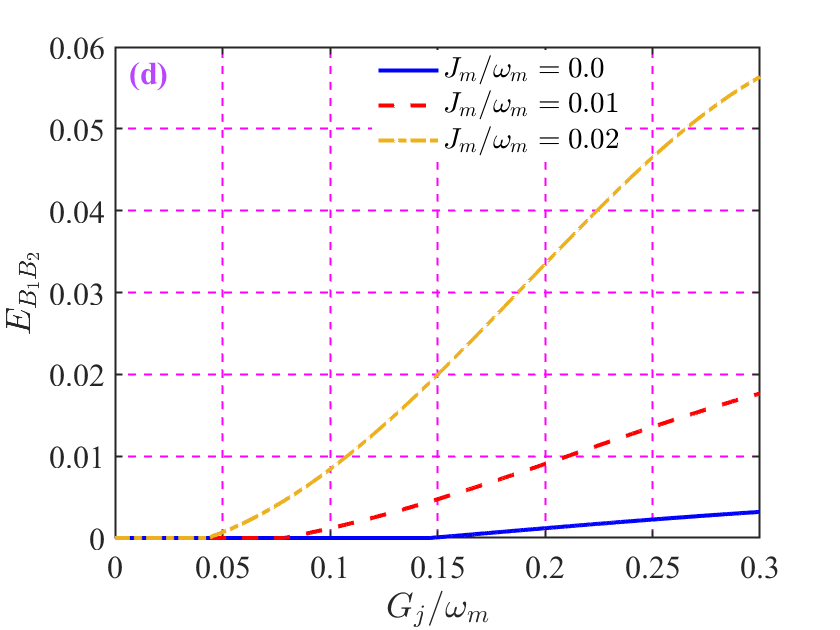}}
	\end{center}
	\caption{(a) Density plot of bipartite $E_{B_1B_2}^N$ vs the scaled detuning $\tilde{\Delta}_a/\omega_m$ and the optomechanical coupling strength $G_j/\omega_m$ when $J_m/\omega_m=0.0$, N=100, $\kappa/\omega_m=1/3$, and $\theta=0$. (b) Density plot of bipartite, $E_{B_1B_2}^N$ vs the scaled detuning $\tilde{\Delta}_a/\omega_m$ and the optomechanical coupling strength $G_j/\omega_m$ when $J_m/\omega_m=0.02$, N=100, $\kappa/\omega_m=1/3$, and $\theta=\pi/2$. (c) Density plot of bipartite entanglement $E_{B_1B_2}^N$ vs the optomechanical coupling strength $G_j/\omega_m$ and couplng constant $J_m/\omega_m$, when $\tilde{\Delta}_a/\omega_m=1.5$, N=100, $\kappa/\omega_m=1/3$, and $\theta=\pi/2$. (d) Bipartite entanglement $E_{B_1B_2}^N$ vs the optomechanical coupling strength $G_j/\omega_m$ for different values of $J_m/\omega_m$, when $\tilde{\Delta}_a/\omega_m=1.5$, N=100, $\kappa/\omega_m=1/3$, and $\theta=\pi/2$. Here,  $\gamma_m/\omega_m$=0.3, M=N/2, and $n_{th}=0.001$. }
	\label{fig:fig4}
\end{figure*}
This section investigates the enhancement of the bipartite entanglement between our two molecular modes $B_1$ and $B_2$ under their inter molecular coupling $J_m$. Based on our dark mode analysis from \autoref{fig:fig2}, we will work under the Dark Mode Breaking regime (\textbf{DMB}) by considering $\theta=\pi/2$. \autoref{fig:fig3} shows the bipartite entanglement between the optical mode and the second molecular mode ($E_{aB_2}^N$, and similar result can be obtained with the first molecular mode $E_{aB_1}^N$, but not shown). 
\begin{figure*}[tbh]
	\begin{center}
		\resizebox{0.45\textwidth}{!}{
			\includegraphics{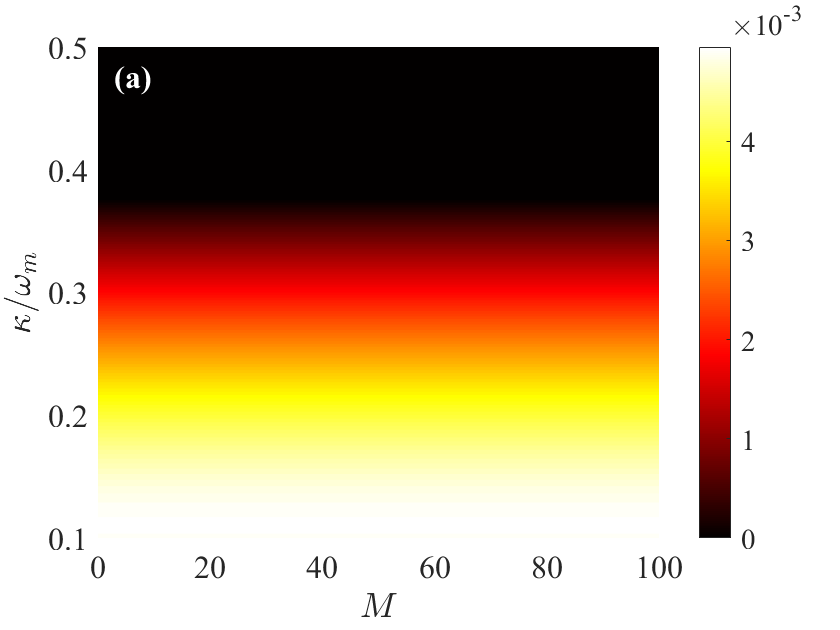}}
		\resizebox{0.45\textwidth}{!}{
			\includegraphics{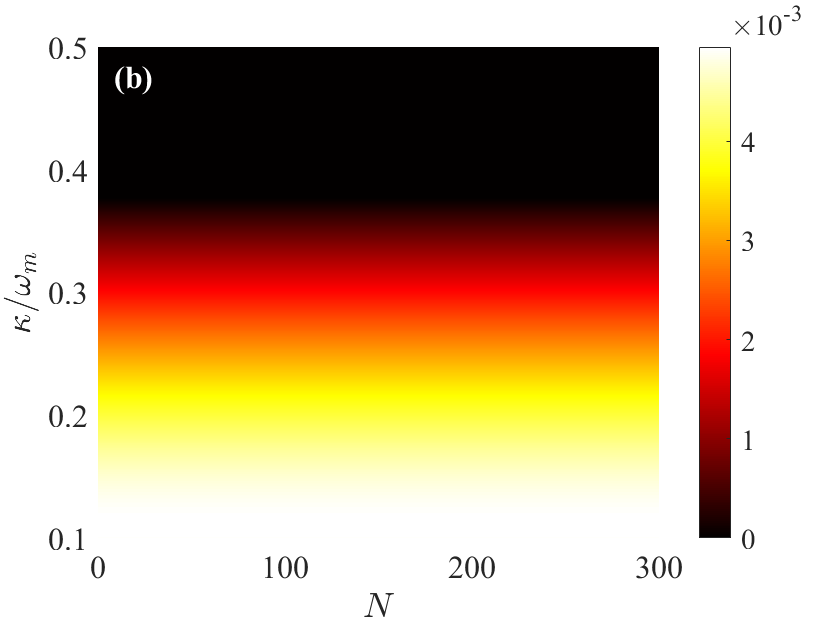}}
		\resizebox{0.45\textwidth}{!}{
			\includegraphics{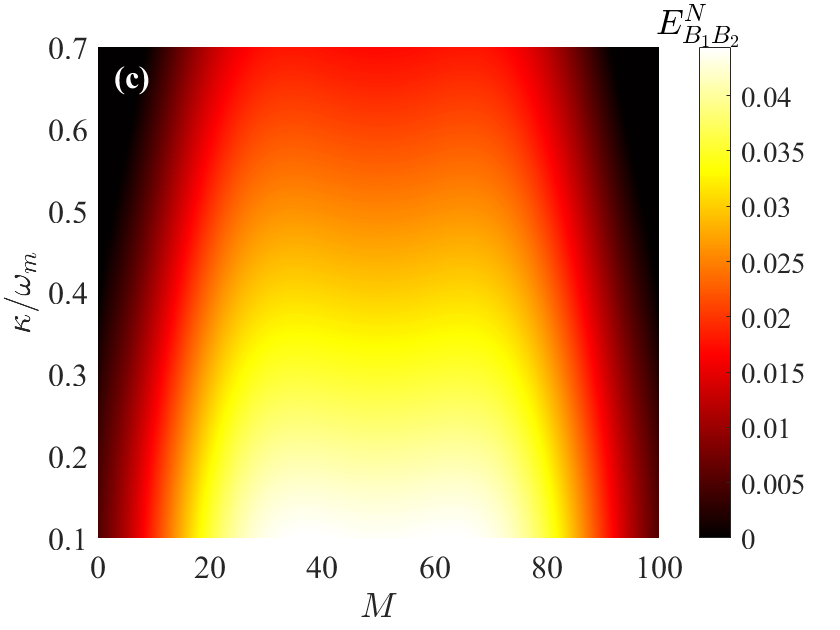}}
		\resizebox{0.45\textwidth}{!}{
			\includegraphics{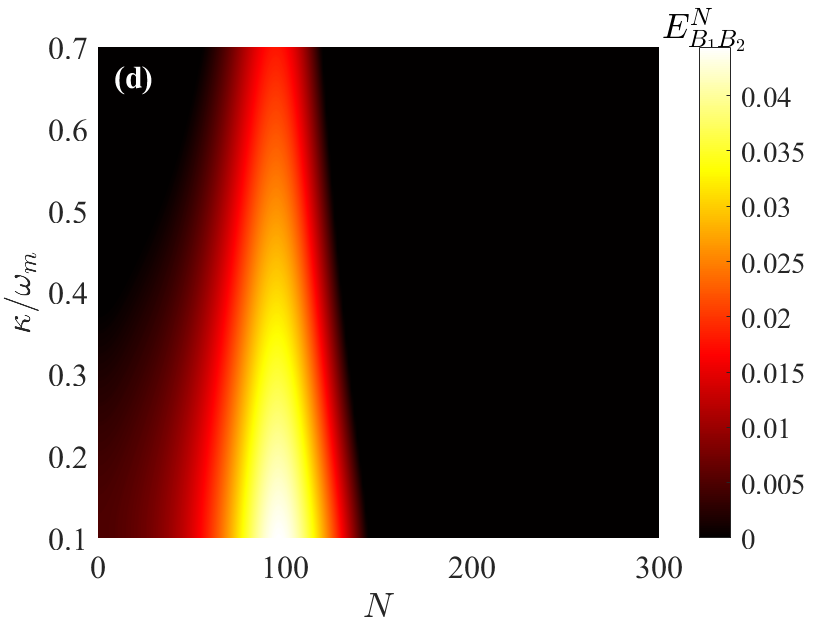}}
	\end{center}
	\caption{(a) Density plot of bipartite entanglement, $E_{B_1B_2}^N$ vs the distribution number of collective mode, M and the scaled decay rate $\kappa/\omega_m$ when $J_m/\omega_m=0$ and $\theta=0$. (b) Density plot of bipartite entanglement, $E_{B_1B_2}^N$ vs the total number of molecules, N, and scaled decay rate, $\kappa/\omega_m$ when M=N/2, $J_m/\omega_m=0$ and $\theta=0$. (c) Density plot of bipartite entanglement, $E_{B_1B_2}^N$ vs the distribution number of collective mode, M and the scaled decay rate $\kappa/\omega_m$ when $J_m/\omega_m=0.02$ and $\theta=\pi/2$. (d) Density plot of bipartite entanglement, $E_{B_1B_2}^N$ vs the total number of molecules, N, and the scaled decay rate, $\kappa/\omega_m$ when M=N/2, $J_m/\omega_m=0.02$, $G_j/\omega_m$=0.2 and $\theta=\pi/2$. Common parameters are chosen as, $\gamma_m/\omega_m$=0.3, $\tilde{\Delta}_a/\omega_m$=1.5, N=100; M=50, and $T= 210$ K. }
	\label{fig:fig5}
\end{figure*}
\autoref{fig:fig3}a depicts the entanglement $E_{aB_2}^N$ versus the effective optomechanical coupling strength $G_j$ and the effective detuning $\tilde{\Delta}_a$ for $J_m=0$, while \autoref{fig:fig3}b displays the $E_{aB_2}^N$ versus the inter molecular coupling $J_m$ and the coupling $G_j$ for $\tilde{\Delta}_a=\omega_m$. It can be seen that entanglement is generated from a certain threshold of the optomechanical coupling ($G_j\gtrsim0.05\omega_m$), and that generated entanglement reaches its optimal value around the resonance  (see \autoref{fig:fig3}a). Another feature to mention is that the dark mode breaking effect does not significantly enhance the entanglement between the optical and molecular modes as displayed in \autoref{fig:fig3}b. In fact, colorbars of \autoref{fig:fig3}a ($J_m=0$) and \autoref{fig:fig3}b ($J_m\neq0$) shows similar optimal values of entanglements. This result can be understood based on the fact that the inter coupling $J_m$ is not directly related to the optical mode, but is rather connected to molecular modes. This means that the dark mode breaking effect is more valuable for the enhancement of the molecular-molecular entanglement or the tripartite entanglement.    

To figure out the effect of the \textbf{DMB} in the enhancement of molecular-molecular entanglement, we displayed the set of \autoref{fig:fig4}. Contour plots of the molecular-molecular entanglement $E_{B_1B_2}^N$ versus the coupling strength $G_j$ and the detuning $\tilde{\Delta}_a$ are represented in \autoref{fig:fig4}a (for $J_m=0$) and in \autoref{fig:fig4}b (for $J_m\neq0$). When $J_m=0$ (see \autoref{fig:fig4}a), it can be seen that the inter molecular entanglement is generated for a relatively strong driving strength ($G_j\sim0.15\omega_m$), while this threshold is low when the coupling $J_m$ is turned on (see \autoref{fig:fig4}b). Moreover, the molecular entanglement is generated out of resonance ($\tilde{\Delta}_a\gtrsim2\omega_m$), while this entanglement is generated close to the resonance ($\tilde{\Delta}_a\sim\omega_m$) in \autoref{fig:fig4}b. Furthermore, one also observe that the parameter area space's for generating molecular entanglement widens as the inter molecular coupling $J_m$ is accounted (compare \autoref{fig:fig4}a and \autoref{fig:fig4}b). Regarding the amount of the generated entanglement, one observes at least a threefold entanglement enhancement under the \textbf{DMB} as we compare the colorbars of \autoref{fig:fig4}a and \autoref{fig:fig4}b. It results that the inter molecular coupling $J_m$ is an interesting physical parameter for the entanglement engineering in our proposal. Indeed, this parameter significantly improve the generated  entanglement for low-threshold driving strength, which is of great importance in experimental investigations. Therefore, the dark mode breaking effect constitutes an interesting tool to control and for enhancing entanglement. To further provide insights on the key role of the coupling $J_m$ regarding the entanglement enhancement, we displayed a contour plot of $E_{B_1B_2}^N$ versus $J_m$ and $G_j$ in \autoref{fig:fig4}c. One observes an enhancement of the entanglement as the coupling $J_m$ increases up to $J_m=0.02\omega_m$, where the entanglement starts to decrease. This reveals that the \textbf{DMB} requiers a strategical control to be efficient for the entanglement enhancement. This feature is further depicted in \autoref{fig:fig4}d, where the entanglement is displayed versus $G_j$ for different values of $J_m$. It can be clearly observed an improvement of the entanglement, and the low threshold driving strength entanglement as the coupling $J_m$ increases.            
\begin{figure*}[tbh]
	\begin{center}
		\resizebox{0.45\textwidth}{!}{
			\includegraphics{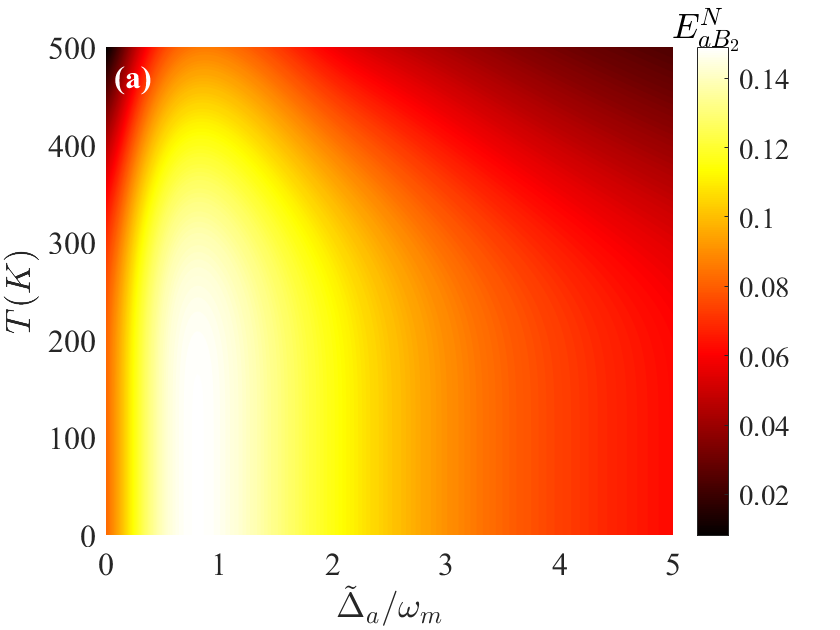}}
		\resizebox{0.45\textwidth}{!}{
			\includegraphics{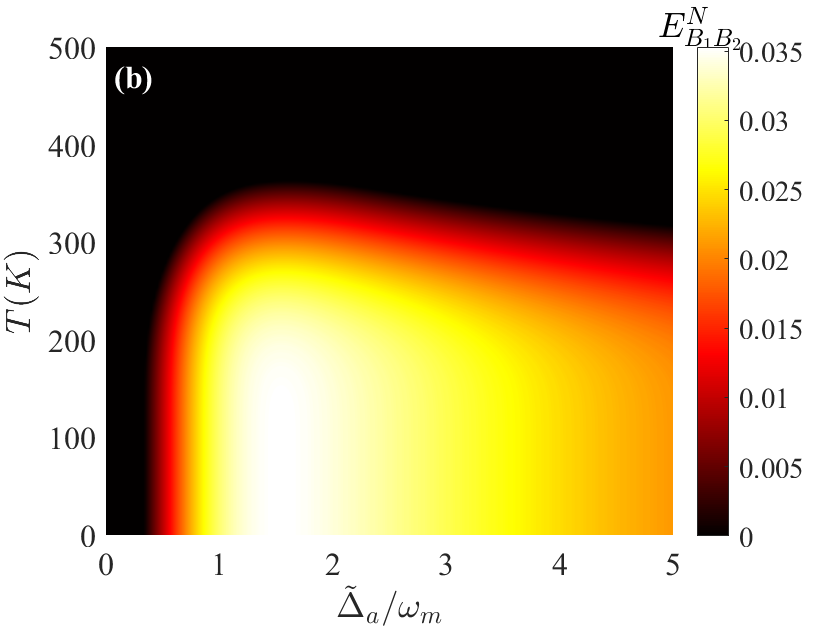}}
		\resizebox{0.45\textwidth}{!}{
			\includegraphics{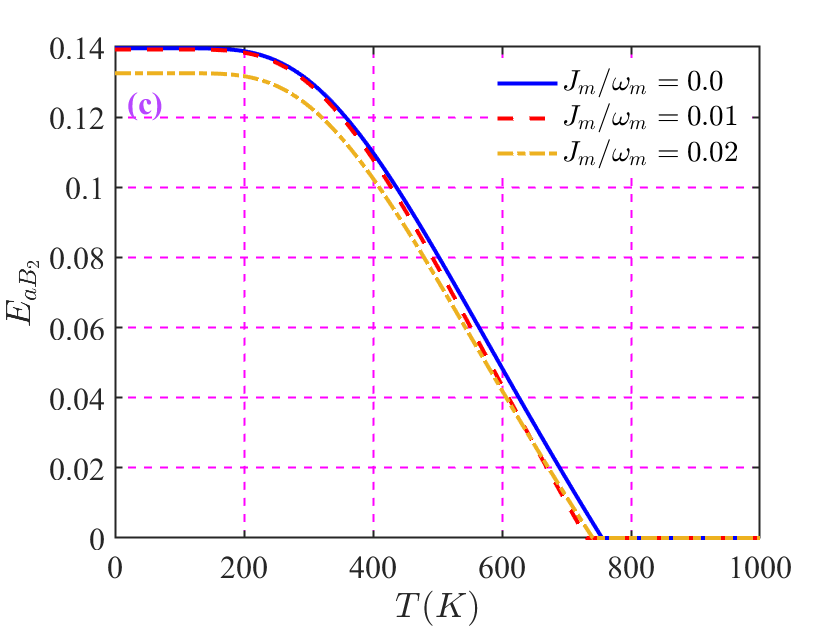}}
		\resizebox{0.45\textwidth}{!}{
			\includegraphics{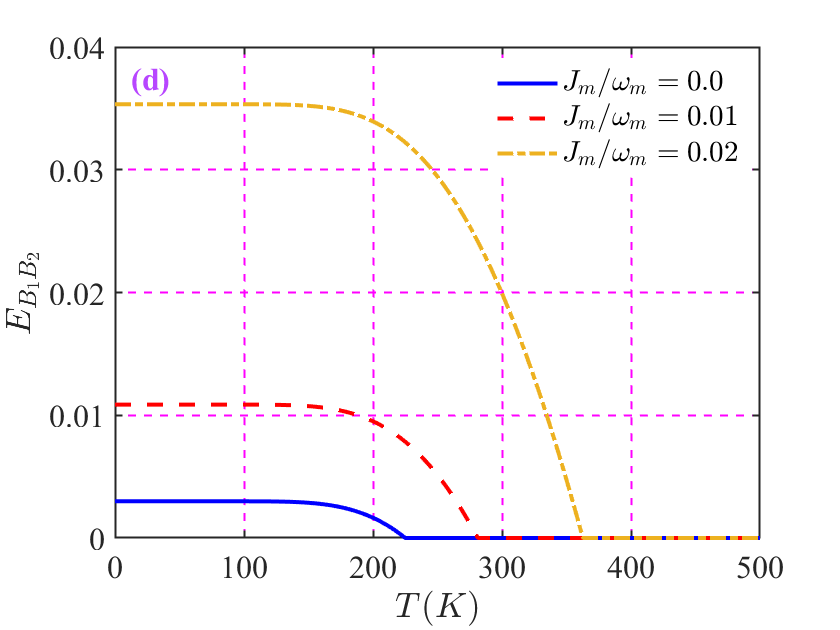}}
	\end{center}
	\caption{ (a) Density plot of bipartite entanglement, $E_{aB_2}^N$ vs $\tilde{\Delta}_a/\omega_m$ and temperature T, when $J_m/\omega_m=0.02$ and $\kappa/\omega_m=1/3$. (b) Density plot of bipartite entanglement, $E_{B_1B_2}^N$ vs $\tilde{\Delta}_a/\omega_m$ and temperature T, when $J_m/\omega_m=0.02$ and $\kappa/\omega_m=1/3$. (c) Plot of bipartite entanglement, $E_{aB_2}^N$ vs  temperature T for different values of $J_m/\omega_m$, when $\tilde{\Delta}_a/\omega_m$=1.5  and $\kappa/\omega_m=1/3$. (d) Plot of bipartite entanglement, $E_{B_1B_2}^N$ vs  temperature T for different values of $J_m/\omega_m$,  when $\tilde{\Delta}_a/\omega_m$=1.5 and $\kappa/\omega_m=1/3$. Here, $G_j/\omega_m=0.2$, $\gamma_m/\omega_m$=0.3, N=100, M=N/2, $\theta=\pi/2$. Other parameters are the same as for \Cref{fig:fig3}.}
	\label{fig:fig6}
\end{figure*}

It is also interesting to study the generated entanglement depending on the molecular numbers in our proposal. Such an investigation is carried out in \autoref{fig:fig5}, where the entanglement $E_{B_1B_2}^N$ is displayed versus the molecular numbers (N and M) and the cavity decay rate $\kappa$. Indeed, \autoref{fig:fig5}a and \autoref{fig:fig5}c represent $E_{B_1B_2}^N$ versus the collective molecular mode M and the cavity decay rate $\kappa$, where $J_m=0$ and $J_m=0.02\omega_m$, respectively. It can be seen that the entanglement is optimaly centered around the distribution number of the collective molecular mode M. Moreover, this entanglement is generated in the sideband resolved regime (or good-cavity regime, i.e., $\kappa/\omega_m<1$), when $J_m=0$ (see \autoref{fig:fig5}a). Under the dark mode breaking effect, \autoref{fig:fig5}c shows how the space suitable for entanglement generation shrinks and the generated entanglement spams over less collective molecular modes compared to \autoref{fig:fig5}a. 
\begin{figure}[tbh]
	\begin{center}
		\resizebox{0.45\textwidth}{!}{
			\includegraphics{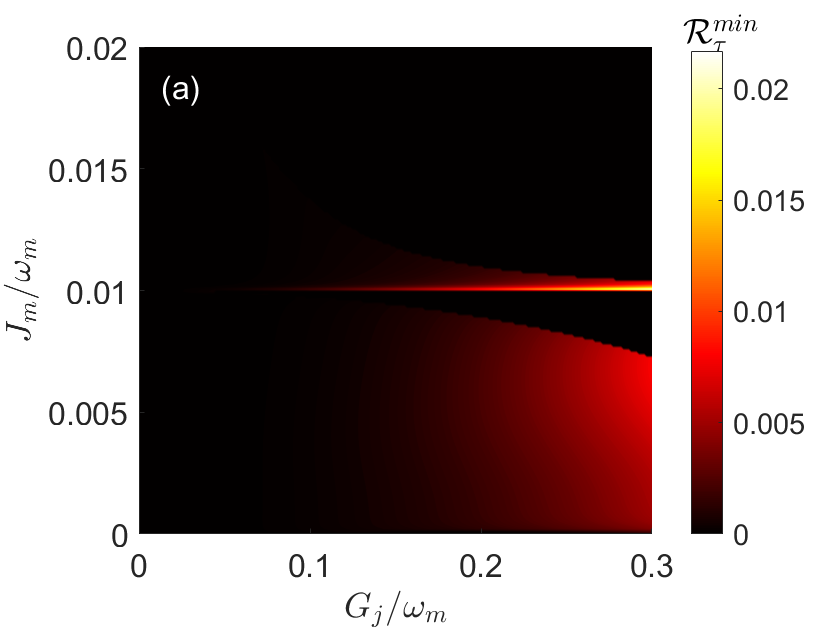}}
		\resizebox{0.45\textwidth}{!}{
			\includegraphics{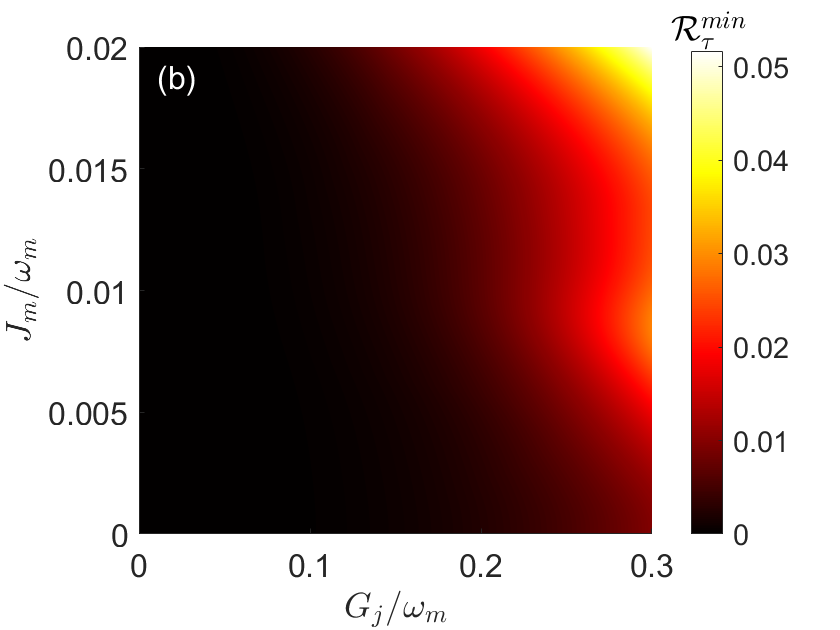}}
	\end{center}
	\caption{(a) Residual cotangle measure $\mathcal{R}^{min}_\tau$ versus the optomechanical coupling strength $G_j$ and coupling strength between the vibrational modes, $J_m$  when $\gamma_m=10^{-3}\omega_m$. (b) Residual cotangle measure $\mathcal{R}^{min}_\tau$ versus the optomechanical coupling strength $G_j$ and $J_m$  when, $\gamma_m=0.3\omega_m$ and $\kappa/\omega_m=0.2$. Common parameters are chosen as, $\tilde{\Delta}_a/\omega_m$=1.5, N=200, M=N/2, $\kappa/\omega_m=0.2$,  $\theta=\pi/2$, and $n_{th}=0.001$ (corresponding to $T\approx 210$ K), and other parameters are the same as those in Fig 2.}
	\label{fig:fig7}
\end{figure}
It can be also seen that the entanglement is generated over the unresolved sideband regime (or bad-cavity regime, i.e., $\kappa/\omega_m\gtrsim1$) More importantly, one observe at least threefold entanglement enhancement under the dark mode breaking control as shown by the colorbars of \autoref{fig:fig5}a and \autoref{fig:fig5}c. Regarding the total molecular number N, \autoref{fig:fig5}b and \autoref{fig:fig5}d lead to qualitatively  similar results as those concluded from  \autoref{fig:fig5}a and \autoref{fig:fig5}c.
\begin{figure}[tbh]
	\begin{center}
		\resizebox{0.45\textwidth}{!}{
			\includegraphics{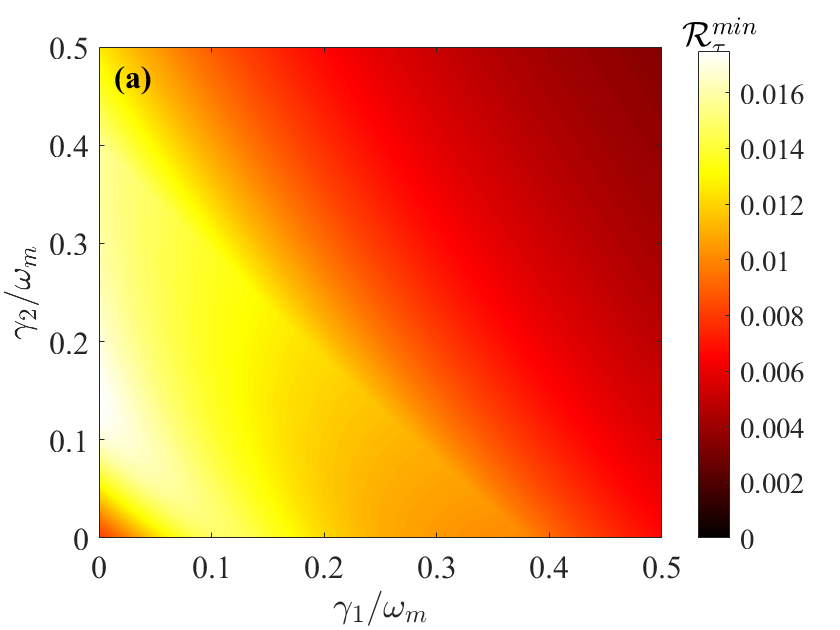}}
		\resizebox{0.45\textwidth}{!}{
			\includegraphics{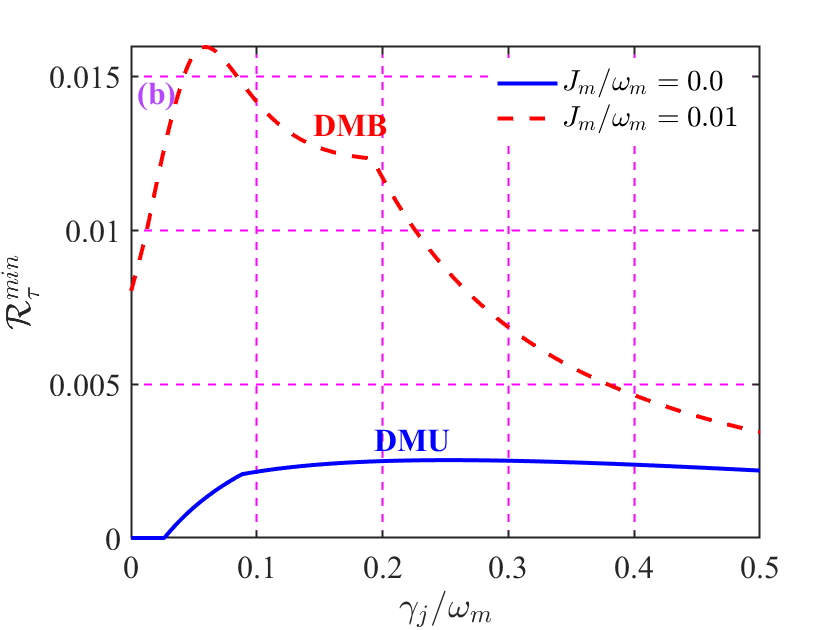}}
	\end{center}
	\caption{(a) Residual cotangle measure $\mathcal{R}^{min}_\tau$  vs molecular decay rates, $\gamma_1$ and $\gamma_2$ in the \textbf{DMB } regime when $J_m=0.01\omega_m$. (b)  Residual cotangle measure vs molecular damping rate $\gamma_m$, extracted from (a). The other parameters are $\kappa/\omega_m=0.2$, $G_j/\omega_m=0.2$, $\tilde{\Delta}_a/\omega_m$=1.5, N=200, M=N/2, $\theta=\pi/2$, and $n_{th}=0.001$ (corresponding to $T\approx 210$ K) and the rest of the parameters are the same as those in \Cref{fig:fig3}.}
	\label{fig:fig8}
\end{figure}
 Indeed, the generated entanglement is limited in  the resolved sideband when $J_m=0$ (\autoref{fig:fig5}b), while this entanglement goes up to the unresolved sideband regime under the dark mode breaking effect $J_m\neq0$ (\autoref{fig:fig5}d). Furthermore, the \textbf{DMB} induces a great improvement of the entanglement, i.e., eith times greather than what is generated in the unbreaking dark mode regime. Such a great enhancement of the entanglement reveals the interesting role of the dark mode control to empower modern quantum technologies based entanglement. More interestingly, generating entanglement in the unresolved sideband regime is an important aspect for experimental investigation, since it relaxes the steadious experimental constraints to work under the resolved sideband regime.   
Another interesting feature to point out about the bipartite entanglements investigated here is the thermal management of these quantum resources. The set of figures displayed in \autoref{fig:fig6} give some insights on the thermal fluctuation on both entanglements $E_{aB_2}^N$ and $E_{B_1B_2}^N$. \autoref{fig:fig6}a represent a contour plot of the entanglement $E_{aB_2}^N$ versus detuning $\tilde{\Delta}_a$ and the bath temperature $T$, while \autoref{fig:fig6}c depicts the same quantity versus the temperature at a fixed detuning $\tilde{\Delta}_a=1.5\omega_m$, for different values of $J_m$. It can be seen that the optimal generated entanglement is around the resonance $\tilde{\Delta}_a\sim1.5\omega_m$, and the entanglement resists to temperature beyond $500 K$ (see \autoref{fig:fig6}a for $J_m=0.02\omega_m$). As aforementioned, breaking the dark mode does not have an impact on the optical-molecular entanglement $E_{aB_2}^N$, owing to the fact the the coupling $J_m$ relies the molecular subsystem. This feature is report on \autoref{fig:fig6}c, where the effect of $J_m$ seems detrimental regarding the robustness of $E_{aB_2}^N$ against temperature. Indeed, as the coupling $J_m$ increases, the entanglement $E_{aB_2}^N$ becomes fragile against bath temperature. The thermal management of the inter molecular entanglement $E_{B_1B_2}^N$ is displayed in \autoref{fig:fig6}b, where the entanglement persists against temperature up to $400K$ under the dark mode breaking effect. Moreover, \autoref{fig:fig6}d shows this inter molecular entanglement versus temperature, at the optimal detuning $\tilde{\Delta}_a=1.5\omega_m$, and for different values of the coupling $J_m$. This figure figures out the effect of the dark mode breaking on the thermal management of  $E_{B_1B_2}^N$. As expected, one observes an improvement of the robustness against temperature as the coupling $J_m$ increases. This result reflects the merit of breaking dark mode in our proposal in order to enhance inter molecular entanglement, which constitutes a crucial quantum resource for quantum technologies including, quantum information processing, quantum computing, quantum chemistry computational tasks. 

\subsection{Enhancement of tripartite entanglement}

In order to benefict from the dark mode breaking effect in our proposal, we need to investigate tripartite entanglement which may unifies the entanglement of the three modes involved in our proposal. 
\begin{figure}[tbh]
	\begin{center}
		\resizebox{0.45\textwidth}{!}{
			\includegraphics{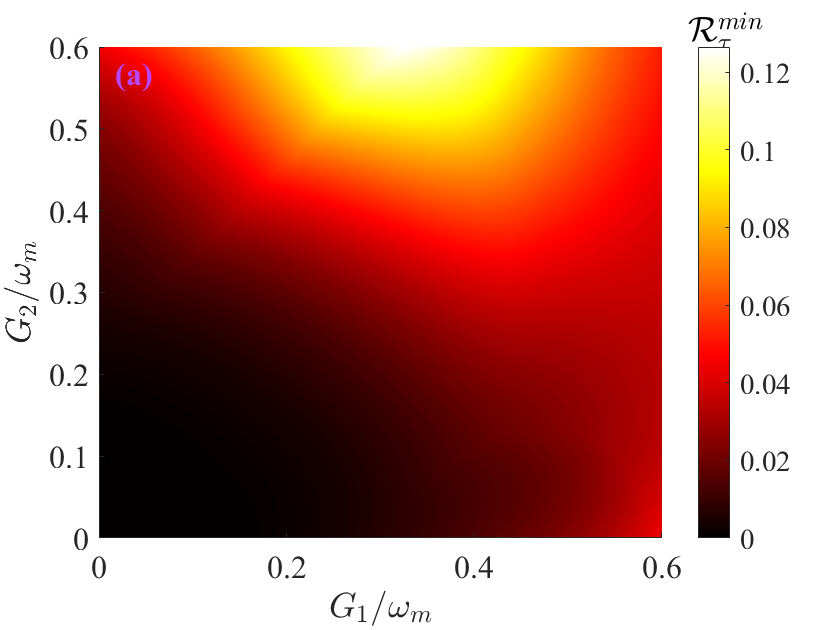}}
		\resizebox{0.45\textwidth}{!}{
			\includegraphics{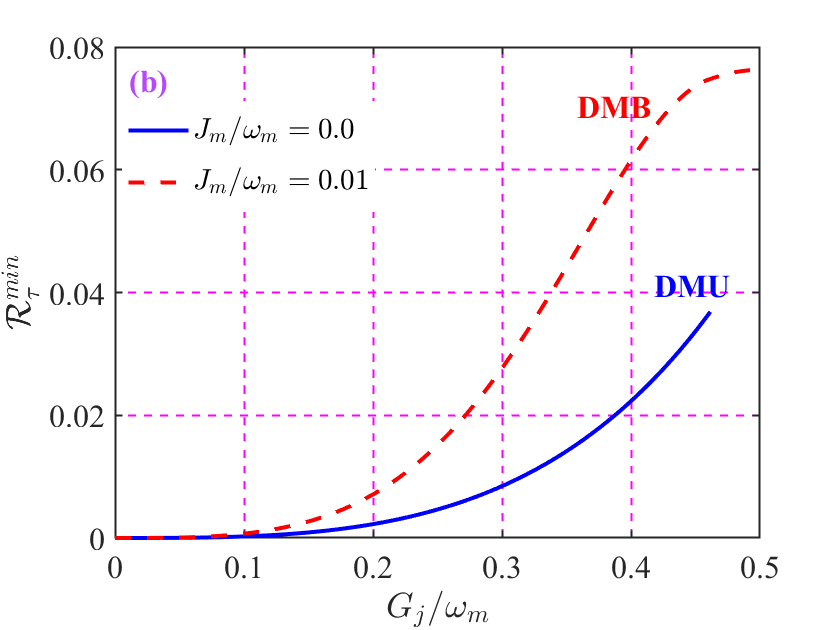}}
	\end{center}
	\caption{(a) Residual cotangle measure $\mathcal{R}^{min}_\tau$ versus the optomechanical coupling strength $G_1$ and $G_2$ in the \textbf{DMB} regime when $\gamma_m/=0.3\omega_m$. (b) Tripartite  entanglement versus optomechanical coupling strength, $G_j/\omega_m$ extracted from (a) for different values of $J_m$, i.e., $J_m=0$ (full line), $J_m=0.01\omega_m$ (dashed line), and $J_m=0.02\omega_m$ (dash-dotted line). The used parameters are, $\tilde{\Delta}_a/\omega_m$=1.5, N=200, M=N/2, $\theta=\pi/2$ and $n_{th}=0.001$ (corresponding to $T\approx 210$ K) when they are not variables and other parameters are the same as those in \Cref{fig:fig3}.}
	\label{fig:fig9}
\end{figure}
For this purpose, we have plotted \autoref{fig:fig7} which displays the contour plot of the residual cotangle versus both couplings $G_j$ and $J_m$, for $\gamma_m=10^{-3}\omega_m$ (\autoref{fig:fig7}a) and $\gamma_m=0.3\omega_m$ (\autoref{fig:fig7}b). It can be seen that tripartite entanglement does not exist (or is very weak) in the unbreaking dark mode regime, especially for $J_m=0$. As the coupling $J_m$ increases, one observes an increase of the cotangle. Moreover, the generation of tripartite entanglement require a certain threshold of the driving strength $G_j$ as depicted in \autoref{fig:fig7}. More importantly, the amount of the generated entanglement is higher in \autoref{fig:fig7}b as compared with \autoref{fig:fig7}a. This reveals that our proposal does not require high quality mechanical factor to produces strong tripartite entanglement. Such feature is interesting for experimental investigations, as it relaxes constraints related to the engineering high quality factor mechanical resonators. 

In order to further get insights about this feature, we have displayed a contour plot of the cotangle versus both mechanical decay rates in \autoref{fig:fig8}. As aforementioned, \autoref{fig:fig8}a shows how tripartite entanglement becomes significant when the mechanical dissipations start to compete with the optical decay rate $\kappa$. Indeed, as the condition $\kappa\sim \gamma_m\approx0.2\omega_m$ is fulfilled ( \autoref{fig:fig8}a), the generated cotangle gets stronger under the dark mode breaking effect. This can be further seen in  \autoref{fig:fig8}b, where we displayed cotangle versus the mechanical decay rates  $\gamma_m$ (considering degenerated mechanical resonators), for different couplings $J_m$. In the dark mode unbreaking regime (see full line in \autoref{fig:fig8}b), the tripartite entanglement slightly grwoths no matter the strength of the mechanical decay rate. However, the entanglement shows great enhancement in the dark mode breaking regime (see dashed line). Moreover, that entanglement enhancement is optimal when the mechanical dissipation ($\gamma_m$) rate compensates the optical decay rate ($\kappa$). From that optimal value, the cotangle starts decreasing and reaches the amount of entanglement generated when the dark mode is unbroken in the deep reversed dissipative regime ($\gamma_m>\kappa$). This feature reveals that the generated entanglement in the reversed dissipative regime is fragile, even though it is high  compared to what is generated for high quality mechanical factor ($\gamma_m \ll\kappa$).          
\begin{figure}[tbh]
	\begin{center}
		\resizebox{0.45\textwidth}{!}{
			\includegraphics{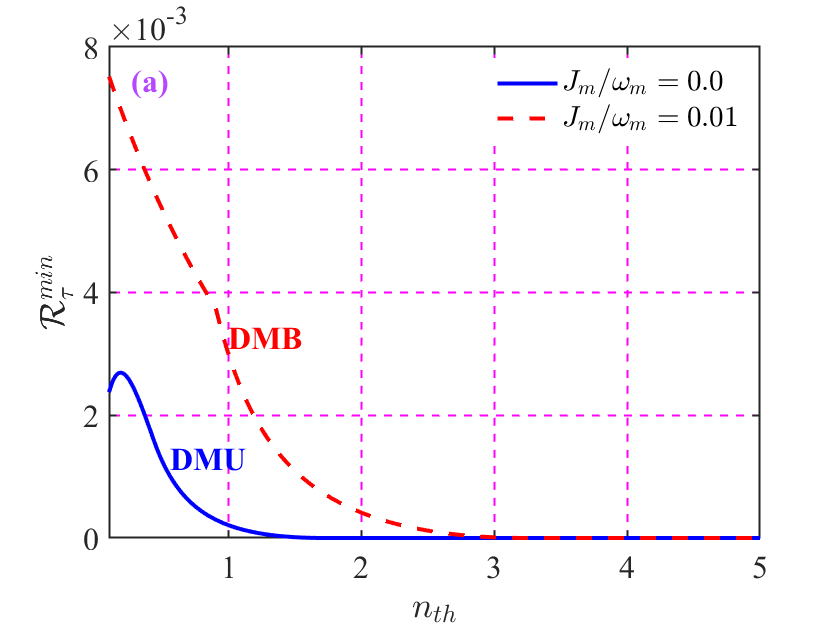}}
		\resizebox{0.45\textwidth}{!}{
			\includegraphics{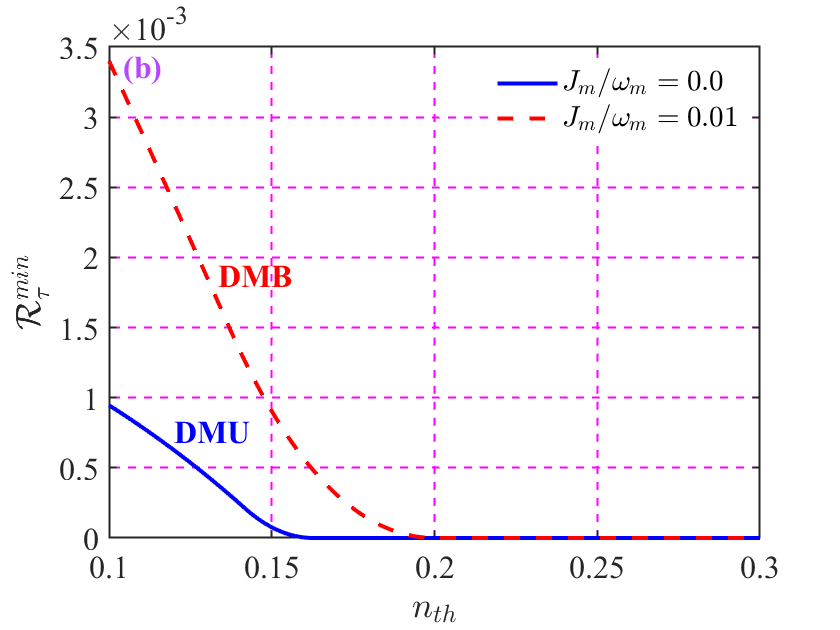}}
	\end{center}
	\caption{(a) Residual cotangle measure $\mathcal{R}^{min}_\tau$ versus thermal phonon number $n_{th}$ in the \textbf{DMU}  and \textbf{DMB} regimes when $\gamma_m=10^{-3}\omega_m$. (b) Residual cotangle measure $\mathcal{R}^{min}_\tau$ versus thermal phonon number $n_{th}$ in the \textbf{DMU}  and \textbf{DMB} regimes when $\gamma_m=0.3\omega_m$. Parameters are chosen as, $\kappa/\omega_m=0.2$, $G_j/\omega_m=0.2$, $\tilde{\Delta}_a/\omega_m=1.5$, N=200, M=N/2, and $\theta=\pi/2$. Other parameters are the same as those in \Cref{fig:fig3}.}
	\label{fig:fig10}
\end{figure}

To further figure out the effect of the driving strength on the cotangle, we displayed a contour plot of tripartite entanglement versus both coupling $G_1$ and $G_2$ in  \autoref{fig:fig9}. It can be seen that the generated entanglement requires a certain threshold of the driving strength ($G_j\sim 0.2 \omega_m$, see \autoref{fig:fig9}a), which is reachable in the molecular optomechanical state-of-the-art experiments. This feature is examplified in \autoref{fig:fig9}b, where the cotangle extracted from \autoref{fig:fig9}a is displayed for different values of $J_m$, i.e., $J_m=0$ (full line), $J_m=0.01\omega_m$ (dashed line), and $J_m=0.02\omega_m$ (dash-dotted line).  It can be observed that there is an significant enhancement of the tripartite entanglement under the dark mode breaking effect. This reveals how breaking dark mode in our proposal is an interesting tool to control entanglement generation.   
Another interesting feature to investigate is the thermal management of the residual cotangle, which quantifies the resilience of the tripartite entanglement against thermal fluctuations. This is carried out through \autoref{fig:fig10}, where the residual cotangle is represented versus the thermal phonon number $n_{th}$ for high quality factor for vibrational modes (\autoref{fig:fig10}a) and in the reversed dissipative regime (\autoref{fig:fig10}b). For the high quality factor of vibrational modes case, one observes at least a threefold robustness enhancement of the entanglement against thermal fluctuation under the dark mode breaking effect, compared with the dark mode unbreaking regime (compare full and dashed lines in \autoref{fig:fig10}a). Once more, this reflects how breaking dark mode can be used as a tool for thermal management in our proposal, leading to a generation of highly thermal resilient quantum resources. These generated noise-tolerant quantum resources are crucial for modern quantum technologies including, quantum information processing, quantum communication, and quantum computational tasks. In the reversed dissipative regime (\autoref{fig:fig10}b), despite the fact that the amount of the generated entanglement is high, it seems to be fragile against thermal fluctuations as aforementioned. This can be understood from the fact that the system has more thermal channels to interact with its bath environment, leading to a fast decaying of the generated entanglement. Therefore, the reversed dissipative regime is not suitable for engineering noise-tolerant quantum resources.         

\section{Conclusion}\label{sec:concl}
In summary, we have provided a molecular cavity optomechanical scheme for generating bipartite and tripartite quantum entanglement through the control of the dark mode via tuning the intermolecular coupling $J_m$. By employing a synthetic magnetism field, our system allows flexible switching between the DMU and DMB regimes. We have shown that entanglement is strongly suppressed in the DMU regime but is significantly enhanced when the DM is broken. More importantly, quantum entanglement generated in our scheme remains robust even at elevated temperatures. These results demonstrate that the intermolecular coupling $J_m$ and DM control provide effective tools for entanglement engineering, highlighting the potential of our scheme for applications in quantum information processing and quantum computing.

\section*{Acknowledgments}
P.D. acknowledges the Iso-Lomso Fellowship at Stellenbosch Institute for Advanced Study (STIAS), Wallenberg Research Centre at Stellenbosch University, Stellenbosch 7600, South Africa, and The Institute for Advanced Study, Wissenschaftskolleg zu Berlin, Wallotstrasse 19, 14193 Berlin, Germany. This work was supported by Princess Nourah
bint Abdulrahman University Researchers Supporting Project number (PNURSP2026R893), Princess Nourah bint Abdulrahman University, Riyadh, Saudi Arabia. The authors are thankful to the Deanship of Graduate Studies and Scientific Research at University of Bisha for supporting this work through the Fast-Track Research Support Program
\section*{Author Contributions}
 E.K.B. and P.D. conceptualized the work and carried out the simulations and analysis. A.N.A.-A and H.A. participated in all the discussions and provided useful methodology and suggestions for the final version of the manuscript. P.D. and A.-H. A.-A. participated in the discussions and supervised the work. All authors participated equally in the writing, discussions, and the preparation of the final version of the manuscript
\section*{Competing Interests} 
All authors declare no competing interests.

\section*{Data Availability}
Relevant data are included in the manuscript and supporting information. Supplementary data are available upon reasonable request.

\bibliography{RefMolecularDarkMode}
\end{document}